\documentclass[aps,prd,11pt,showpacs,superscriptaddress,amssymb,amsmath,nofootinbib]{revtex4-1}

\usepackage{amsmath,amssymb}
\usepackage{graphicx}
\usepackage{xcolor}
\usepackage{multirow,booktabs}
\usepackage{tabularx}
\usepackage{array}
\usepackage{rotating}
\usepackage{hyperref}

\begin{document}

\title{The Impact of Spin Priors on Parameterized Tests of General Relativity}

\author{Tian-hao Wu}\email{1020241999@glut.edu.cn}

\author{Shu-Jun Rong}
\email{rongshj@glut.edu.cn}

\affiliation{
College of Physics and Electronic Information Engineering,
Guilin University of Technology,
Guilin, Guangxi 541004, China
}

\begin{abstract}
Spin priors play a fundamental role in gravitational-wave parameter estimation, yet their impact on parameterized tests of General Relativity (GR) remains insufficiently understood. We systematically investigate how spin-prior choices affect the 1.5PN deviation parameter $\delta \hat{\phi}_3$ using real gravitational-wave events. Prior-induced effects are quantified through the Jensen--Shannon divergence and posterior median shifts. We find that $\chi_p$ is significantly more sensitive to the spin prior than $\chi_{\rm eff}$, while $\delta \hat{\phi}_3$ is generally robust but can exhibit larger shifts in low-information regimes. GW231123\_135430 shows particularly strong sensitivity, but its inferred properties are subject to waveform-model and event-specific systematics, warranting caution in interpreting results strongly influenced by this event. When $\delta \hat{\phi}_3$ is included in the parameter-estimation model, a strong correlation emerges between the prior sensitivities of $\chi_{\rm eff}$ and $\delta \hat{\phi}_3$, consistent with their partial degeneracy at the 1.5PN order, whereas $\chi_p$ remains weakly correlated. A leave-one-out test shows that GW231123\_135430 significantly affects the correlation strength, but the correlation persists when this event is removed. These results demonstrate that spin-prior choices can propagate into parameterized tests of GR and should be carefully accounted for when interpreting apparent deviations from GR.
\end{abstract}

\maketitle

\newpage

\section{Introduction}

Since the first detection of GW150914\cite{LIGOScientific:2016aoc,LIGOScientific:2016vlm,LIGOScientific:2016emj,LIGOScientific:2018mvr}, gravitational wave observations have enabled stringent tests of General Relativity in the strong field and highly dynamical regime\cite{Li:2012,LIGOScientific:2016lio,Isi:2019aib}. A widely used approach is the parameterized post-Einsteinian (PPE) framework\cite{Yunes:2009ke,Battista:2021rlh,Battista:2022hmv,Battista:2022sci,Battista:2023znv,DeFalco:2023djo,DeFalco:2024ojf}. A key advantage of this framework is its model independence, as it does not rely on any specific alternative theory of gravity. This framework can be expressed as:
\begin{equation}
\tilde{h}(f)=
\left\{
\begin{array}{ll}
\tilde{h}^{(\mathrm{GR})}_{\mathrm{I}}(f)\,\cdot\,(1+\alpha u^a)\,e^{i\beta u^b},
& f<f_{\mathrm{IM}}, \\[6pt]
\gamma u^c e^{i(\delta+\epsilon u)},
& f_{\mathrm{IM}}<f<f_{\mathrm{MRD}}, \\[6pt]
\zeta \dfrac{\tau}{1+4\pi^2\tau^2\kappa (f-f_{\mathrm{RD}})^d},
& f>f_{\mathrm{MRD}}.
\end{array}
\right.
\end{equation}
Here $\tilde{h}(f)$ denotes the parametrized gravitational-wave waveform in the frequency domain, and $u\equiv\pi\mathcal{M}f$ is the dimensionless frequency parameter, where $\mathcal{M}=M\eta^{3/5}$ is the chirp mass, $M=m_1+m_2$ is the total mass, and $\eta=m_1m_2/M^2$ is the symmetric mass ratio.

The full waveform is divided into three parts according to the frequency range~\cite{Ajith:2007qp,Khan:2015jqa}: when $f<f_{\mathrm{IM}}$, it corresponds to the inspiral phase; when $f_{\mathrm{IM}}<f<f_{\mathrm{MRD}}$, it corresponds to the merger phase; and when $f>f_{\mathrm{MRD}}$, it corresponds to the ringdown phase. Here, $f_{\mathrm{IM}}$ and $f_{\mathrm{MRD}}$ represent the transition frequencies between the inspiral and merger phases, and between the merger and ringdown phases, respectively.

In the inspiral phase, the waveform can be accurately described by a post-Newtonian (PN) expansion~\cite{Blanchet:2013haa,Maggiore:2007,Buonanno:1998gg}, where the phase is expressed as a series in powers of $(v/c)$, with  $v$ being the characteristic orbital velocity of the binary. Deviations from General Relativity can be introduced by modifying the PN phase coefficients~\cite{Yunes:2009ke,Yunes:2013dva}, which encode different physical effects such as point-mass dynamics and spin contributions~\cite{Arun:2004hn}. In this framework, the parameters $\alpha$ and $\beta$ control the magnitude of amplitude and phase deviations, while $a$ and $b$ determine their frequency dependence through power-law indices~\cite{Yunes:2009ke,Yunes:2013dva}.

In the merger and ringdown phases, the parameters $\gamma, \delta, \epsilon, \zeta, \tau, \kappa,$ and $d$ are used to characterize the phenomenological modifications of the waveform in these regimes, where analytical PN descriptions are no longer sufficient. Through this piecewise parameterization, the validity of General Relativity can be tested separately in different stages of the coalescence\cite{LIGOScientific:2026fcf}.

However, previous studies have shown that, in regimes with weak spin information, commonly adopted uniform spin priors tend to drive the posterior distributions of spin parameters towards low values, particularly near zero\cite{LIGOScientific:2025pvj,Kobayashi:2026lac,Prasad:2026idz}. When alternative spin priors are adopted, the posterior distributions of spin parameters, which were previously concentrated near zero, exhibit significant changes.

Spin effects enter the gravitational-wave waveform through spin--orbit and spin--spin couplings, contributing to higher-order corrections in the phase at post-Newtonian (PN) orders beyond the leading point-mass terms\cite{Blanchet:2013haa,Prasad:2026idz}. In gravitational-wave analyses, the spin effects of binary black holes are commonly characterized by the effective inspiral spin parameter $\chi_{\mathrm{eff}}$ and the effective precession spin parameter $\chi_p$. They are defined as\cite{Racine:2008qv,Hannam:2013oca}:
\begin{equation}
\chi_{\mathrm{eff}} = \frac{m_1 \vec{\chi}_1 \cdot \hat{L} + m_2 \vec{\chi}_2 \cdot \hat{L}}{m_1 + m_2},
\end{equation}
\begin{equation}
\chi_p = \max \left( \chi_{1\perp}, \frac{4q + 3}{4 + 3q} \, q \, \chi_{2\perp} \right).
\end{equation}
The effective inspiral spin parameter $\chi_{\mathrm{eff}}$ is defined as the mass-weighted projection of the component spins along the orbital angular momentum $\hat{L}$. It is the dominant spin parameter entering the waveform phase evolution at leading post-Newtonian order and remains approximately conserved during the inspiral\cite{Seymour:2026bjg}.

The effective precession spin parameter $\chi_p$ characterizes the in-plane spin components that drive orbital precession~\cite{Hannam:2013oca,Schmidt:2012rh}, leading to characteristic amplitude and phase modulations in the observed gravitational-wave signal that affect parameter estimation~\cite{Petersen:2014exq}. Here, $\chi_{i\perp}$ denotes the magnitude of the component of the dimensionless spin vector $\vec{\chi}_i$ perpendicular to the orbital angular momentum, and $q = m_2/m_1 \leq 1$ is the mass ratio.

The definition of $\chi_p$ captures the dominant contribution to precession effects in the binary dynamics by reducing the multi-dimensional spin configuration into an effective single parameter~\cite{Hannam:2013oca}. These precession effects are generally more weakly constrained than the aligned-spin effects characterized by $\chi_{\rm eff}$, due to their subdominant influence on the waveform phase evolution and their dependence on higher-order modulations~\cite{Petersen:2014exq}.

In particular, the dominant spin contribution appears at the 1.5PN order, corresponding to $(v/c)^3$~\cite{Schafer:2018jfw}. As a result, prior-induced changes in the spin posterior can propagate into the inference of the 1.5PN deviation parameter, indicating that parameterized tests of General Relativity may also be sensitive to the choice of spin prior.

Characterizing such prior-induced effects is crucial for assessing the robustness of parameterized tests of General Relativity, as the inferred deviation parameters may depend on prior assumptions. However, the extent to which such prior sensitivity propagates into PN-deviation parameters remains largely unexplored in existing studies~\cite{LIGOScientific:2019fpa,Seymour:2026bjg}.

In this work, we perform a systematic investigation of prior-induced effects on the 1.5PN deviation parameter within the inspiral regime. We investigate this effect using real gravitational-wave events and employ the Jensen--Shannon divergence (JSD) as a quantitative measure of the resulting posterior differences~\cite{Lin:1991zzm,Endres:2003}. The remainder of this paper is organized as follows. We first introduce the methods and analysis tools used in this work, then present the main results along with relevant figures, and finally summarize our findings and conclusions.

\section{Formulation}
\subsection{Spin Priors}

The traditional prior for spin magnitudes is assumed to be uniform, i.e.,~\cite{Prasad:2026idz}
\begin{equation}
p(\chi_i)=1, \quad \chi_i \in [0,1], \quad i=1,2,
\end{equation}
where $\chi_i$ denotes the dimensionless spin magnitudes of the two black holes. The normalization condition is given be~\cite{Prasad:2026idz}
\begin{equation}
\int_{0}^{1} p(\chi_i)\, d\chi_i = 1, \quad i=1,2.
\end{equation}
The commonly adopted prior is uniform in the spin magnitude $\chi$.
However, uniformity in $\chi$ does not correspond to uniformity in the
three-dimensional spin-vector space $\mathcal{V}$. To investigate the
impact of this prior choice, we compare it with an alternative prior
proposed in Ref.~\cite{Prasad:2026idz}, in which the spin vectors are
uniformly distributed within the unit sphere.

Assuming isotropy, this prior corresponds to a probability density that is uniform in the three-dimensional volume. In spherical coordinates $(\chi,\theta,\phi)$, the distribution can be factorized into radial and angular components, yielding~\cite{Prasad:2026idz}
\begin{equation}
p_{\mathcal{V}}(\chi) = 3\chi^2, \quad
p(\theta) = \frac{1}{2}\sin\theta, \quad
p(\phi) = \frac{1}{2\pi},
\end{equation}
where the factor $\chi^2$ arises from the Jacobian of the spherical coordinate transformation.
However, a prior that is uniform in $\chi$ induces a non-uniform distribution in the full three-dimensional configuration space $\mathcal{V}$. Due to the radial volume element in $\mathcal{V}$, the corresponding transformation impliese~\cite{Prasad:2026idz}
\begin{equation}
p_{\mathcal{V}}(\chi) \propto \frac{1}{\chi^2}.
\end{equation}
This distribution increases toward $\chi=0$, corresponding to a greater
prior density for low-spin configurations in the three-dimensional
spin-vector space~\cite{Prasad:2026idz}
\subsection{Bayesian inference}

The central goal of Bayesian inference is to construct the posterior distribution of the model parameters $\theta$ given the observed data $d$, denoted as $p(\theta|d)$~\cite{Thrane:2018qnx,Veitch:2014wba}. This posterior represents the probability density of $\theta$ conditioned on the data and satisfies the normalization condition
\begin{equation}
\int d\theta \, p(\theta|d) = 1.
\end{equation}
According to Bayes' theorem, the posterior distribution is expressed as
\begin{equation}
p(\theta|d) = \frac{\mathcal{L}(d|\theta)\,\pi(\theta)}{Z},
\label{bayes}
\end{equation}
where $\mathcal{L}(d|\theta)$ is the likelihood function, $\pi(\theta)$ is the prior distribution, and $Z$ is the Bayesian evidence given by
\begin{equation}
Z = \int d\theta \, \mathcal{L}(d|\theta)\,\pi(\theta),
\end{equation}
which plays a central role in model selection problems.

In gravitational-wave data analysis, the likelihood function is typically constructed under the assumption of stationary, Gaussian detector noise~\cite{Abbott:2020niy,Veitch:2014wba}. Under this assumption, the likelihood can be written as
\begin{equation}
\mathcal{L}(d|\theta)
\propto
\exp\left[
-\frac{1}{2}
\left\langle
d-h(\theta)
\,\middle|\,
d-h(\theta)
\right\rangle
\right],
\end{equation}
where $h(\theta)$ denotes the gravitational-wave signal predicted by the model, and the noise-weighted inner product is defined as
\begin{equation}
\langle a|b\rangle
=
4\,\mathrm{Re}
\int_{f_{\min}}^{f_{\max}}
\frac{\tilde{a}(f)\tilde{b}^{*}(f)}
{S_n(f)}
\,df,
\end{equation}
where $\tilde{a}(f)$, $\tilde{b}(f)$ denote the Fourier transforms of $a(t)$, $b(t)$, the asterisk denotes complex conjugation, $S_n(f)$ is the one-sided noise power spectral density~\cite{Veitch:2014wba}, and $f_{\min}$, $f_{\max}$ are the lower and upper frequency cutoffs of the analysis.

These expressions make explicit that the posterior distribution depends on both the likelihood and the prior. Consequently, any modification of the prior $\pi(\theta)$ propagates to the posterior $p(\theta|d)$ through Eq.~\eqref{bayes}, particularly when the likelihood provides limited constraints on certain parameters~\cite{Prasad:2026idz}. In such cases, the posterior may inherit features from the prior, potentially introducing biases in inferred quantities. This effect is especially relevant in high-dimensional parameter spaces and in low signal-to-noise ratio regimes, where parameter degeneracies further amplify prior dependence~\cite{Veitch:2014wba,Prasad:2026idz}.
\subsection{Jensen--Shannon divergence}

In this work, we use the Jensen--Shannon divergence (JSD) to quantify the difference between two posterior distributions~\cite{Endres:2003,Fuglede:2004}. The JSD is a symmetric and bounded measure derived from information theory and is widely used to compare probability distributions.

For two distributions $p_1(x)$ and $p_2(x)$, the equal-weight JSD is defined as
\begin{equation}
\mathrm{JSD}(p_1,p_2)=H\!\left(\frac{p_1+p_2}{2}\right)-\frac{1}{2}H(p_1)-\frac{1}{2}H(p_2),
\end{equation}
where
\begin{equation}
H(p)=-\sum_x p(x)\log_2 p(x)
\end{equation}
is the Shannon entropy for a discrete distribution~\cite{Shannon:1948}, with the logarithm taken in base 2.

Since the Shannon entropy is concave, the JSD satisfies
\begin{equation}
\mathrm{JSD}(p_1,p_2)\ge 0,
\end{equation}
with equality if and only if $p_1=p_2$. With this choice of logarithmic base, the JSD is bounded as
\begin{equation}
0 \le \mathrm{JSD}(p_1,p_2) \le 1,
\end{equation}
which allows for a normalized and interpretable measure of the discrepancy. In this paper, a larger JSD indicates a greater difference between the posterior distributions obtained under different prior assumptions.

\section{Results}
\subsection{Prior sensitivity of spin parameters and $\delta \hat{\phi}_3$}
We select 18 gravitational-wave events with signal-to-noise ratio (SNR) greater than 17. The data used in this work are obtained from the Gravitational Wave Open Science Center (GWOSC)\footnote{\url{https://www.gw-openscience.org}}. Based on their original configurations, we perform four independent parameter estimation runs for each event:

(i) reproducing the original analysis following the configuration provided on Zenodo \footnote{\url{https://zenodo.org}}, which adopts the standard uniform spin prior (hereafter referred to as the baseline run);

(ii) using the modified spin prior based on a 3-D topological ball distribution (hereafter referred to as the modified prior);

(iii) using the standard uniform spin prior with the inclusion of the deviation parameter $\delta \hat{\phi}_3$;

(iv) using the modified spin prior with the inclusion of the deviation parameter $\delta \hat{\phi}_3$.

All analyses are performed using the \texttt{bilby} framework (v2.7.1) together with \texttt{bilby\_pipe} (v1.7.0) for parameter estimation~\cite{Ashton:2018jfp,
Romero-Shaw:2020owr}. Tests of general relativity are implemented via \texttt{bilby\_tgr} (v0.2). Waveform generation and detector response are handled by \texttt{lalsuite} (v7.26.1), and posterior sampling is carried out using the nested sampler \texttt{dynesty} (v3.0.0)~\cite{Speagle:2020,
Wette:2020air}. The waveform model employed in this work is the precessing waveform approximant \texttt{IMRPhenomXPHM}~\cite{Pratten:2021,Pratten:2020fqn,Garcia-Quiros:2020qpx,Roy:2025gzv,LIGOScientific:2026oim}.

\begin{figure}
    \centering
    \includegraphics[width=\textwidth]{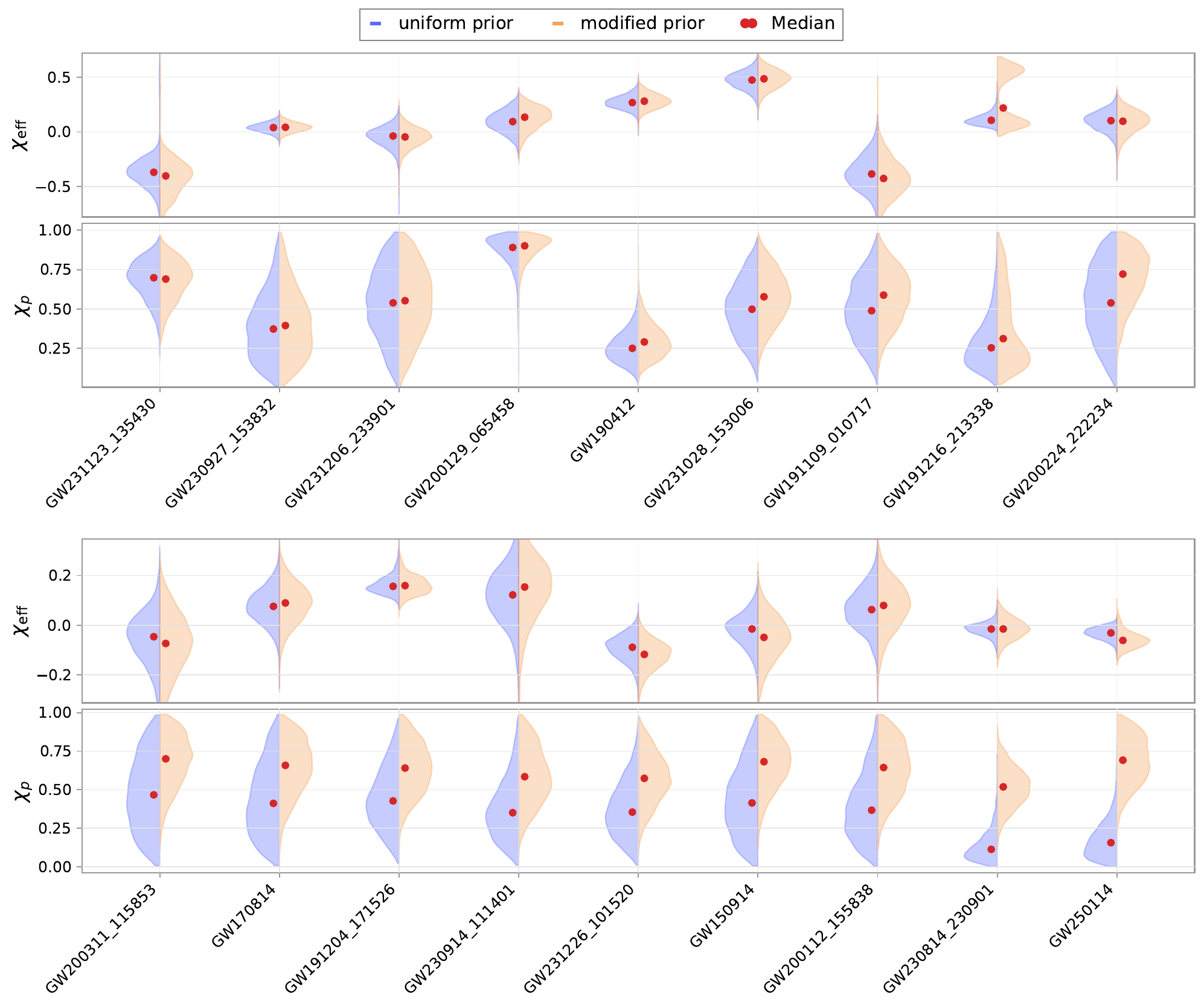}
   \caption{Impact of spin priors on $\chi_{\rm eff}$ and $\chi_p$. $\chi_{\rm eff}$ remains relatively robust, while $\chi_p$ is more sensitive to the prior and shows a noticeable increase in its median.}
    \label{fig:spin}

\end{figure}

Using publicly available posterior samples and posterior samples obtained from newly performed parameter estimation runs, we obtain Fig.~\ref{fig:spin}. As shown in Fig.~\ref{fig:spin}, the violin plots compare the posterior distributions under two spin priors. We find that $\chi_{\rm eff}$ remains relatively stable (see median deviations ), likely due to stronger constraints from the binary mass parameters and the inspiral phasing, whereas $\chi_p$ exhibits a more pronounced change under different prior assumptions, consistent with the generally weaker measurability of precession effects in gravitational-wave observations~\cite{Petersen:2014exq,Vitale:2017}.

\begin{figure}
    \centering
    \includegraphics[width=\textwidth]{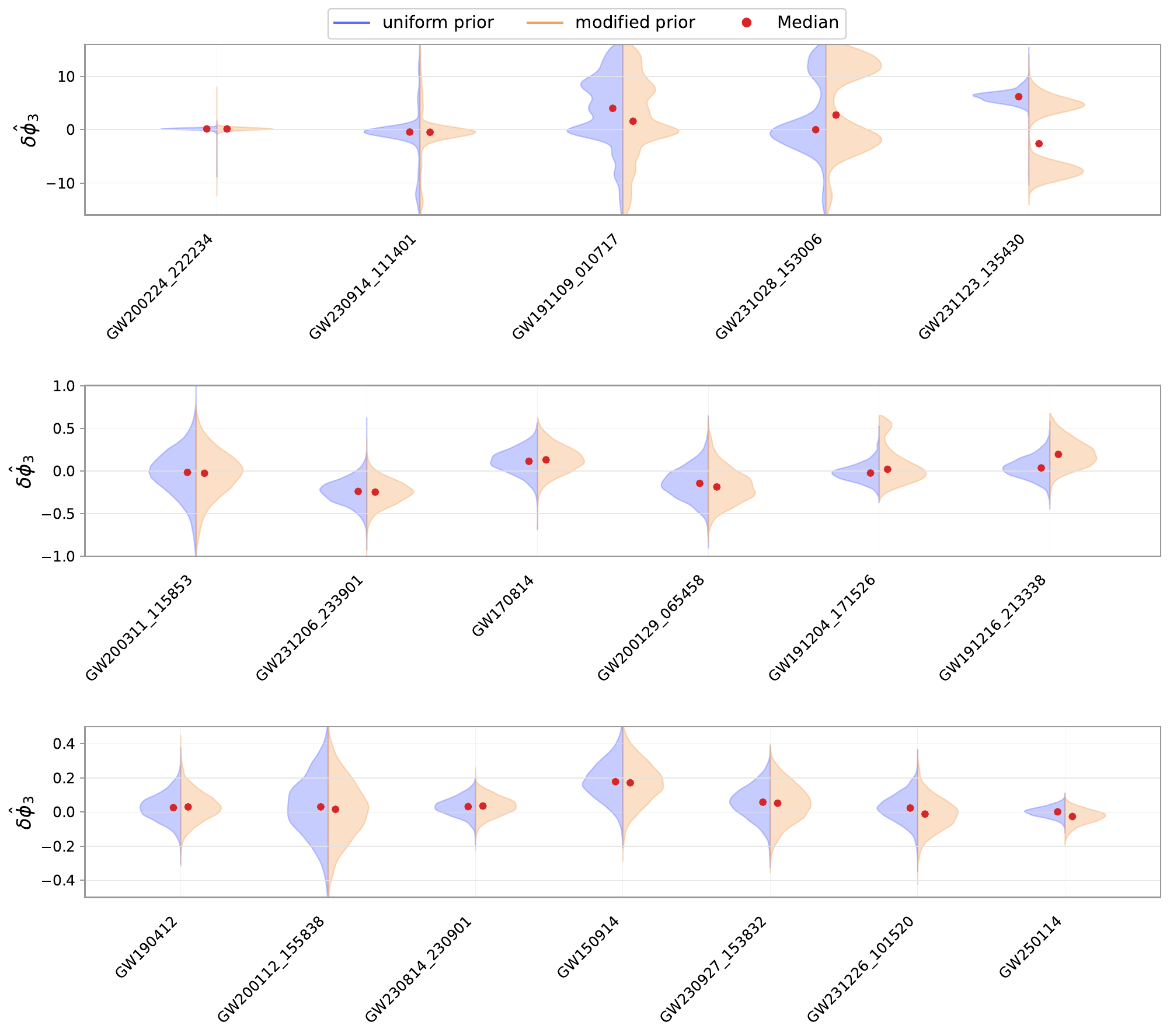}
   \caption{Posterior distributions of the deviation parameter $\delta \hat{\phi}_3$ under uniform and modified spin priors for selected events. The violin plots show the posterior distributions, with red points indicating the median values. Due to the large variation in scale across different events, three representative ranges ($\pm 16$, $\pm 1$, and $\pm 0.5$) are adopted for visualization.}
    \label{fig:dchi3}
 \end{figure}

Employing the same procedure, we construct the comparison plots including the deviation parameter $\delta \hat{\phi}_3$, as shown in Fig.~\ref{fig:dchi3}. Due to the large variation in scale across different events, we adopt three representative ranges for visualization: $\pm 16$, $\pm 1$, and $\pm 0.5$. The large variation in scale is primarily driven by the effective duration of the inspiral phase observed by the gravitational-wave detectors. As the 1.5PN correction predominantly affects the inspiral regime, a short observable inspiral duration provides limited information, leading to a broad and weakly constrained posterior for the deviation parameter $\delta \hat{\phi}_3$~\cite{Seymour:2026bjg,Piarulli:2025lisa}. Conversely, events with a sufficiently long inspiral phase accumulate more signal information, allowing $\delta \hat{\phi}_3$ to be more tightly constrained. We further observe that the largest shifts in the posterior median of $\delta \hat{\phi}_3$ occur in events where the parameter is weakly constrained, such as GW191109\_010717, and GW231028\_153006. In contrast, events with stronger constraints exhibit only mild deviations, indicating that the impact of the prior is amplified in low-information regimes. This behavior is also consistent with the expected degeneracy between spin contributions and PN-deviation parameters at similar orders in the waveform phasing~\cite{Seymour:2026bjg,Narikawa:2016deg}.

The results reporting posterior medians and their symmetric 90\% credible intervals are summarized in Table~\ref{tab:nongr-posterior-summary}.
To characterize changes in the posterior distributions, we employ the Jensen--Shannon divergence (JSD) to quantify relative differences.
We further use the absolute median shift which is $|\Delta \mathrm{med}|$, to quantify the displacement of the posterior median.
\begin{table*}[t]
\centering
\large
\renewcommand{\arraystretch}{2}
\setlength{\tabcolsep}{3.2pt}
\caption{Posterior medians, symmetric 90\% credible-interval offsets, Jensen--Shannon divergence (JSD), and absolute median shifts for the two spin-prior choices.}
\label{tab:nongr-posterior-summary}
\resizebox{\textwidth}{!}{%
\begin{tabular}{l|cccc|cccc|cccc}
\hline
\multirow{2}{*}{{\fontsize{20}{22}\selectfont\textbf{Events}}}
& \multicolumn{4}{c|}{{\fontsize{18}{18}\selectfont$\chi_p$}}
& \multicolumn{4}{c|}{{\fontsize{18}{18}\selectfont$\chi_{\rm eff}$}}
& \multicolumn{4}{c}{{\fontsize{18}{18}\selectfont$\delta \hat{\phi}_3$}} \\
\cline{2-13}
& {\fontsize{13}{16}\selectfont uniform prior}
& {\fontsize{13}{16}\selectfont modified prior}
& {\fontsize{13}{16}\selectfont JSD}
& {\fontsize{13}{16}\selectfont $|\Delta {\rm med}|$}
& {\fontsize{13}{16}\selectfont uniform prior}
& {\fontsize{13}{16}\selectfont modified prior}
& {\fontsize{13}{16}\selectfont JSD}
& {\fontsize{13}{16}\selectfont $|\Delta {\rm med}|$}
& {\fontsize{13}{16}\selectfont uniform prior}
& {\fontsize{13}{16}\selectfont modified prior}
& {\fontsize{13}{16}\selectfont JSD}
& {\fontsize{13}{16}\selectfont $|\Delta {\rm med}|$} \\
\hline
\midrule
GW231123\_135430 & $0.727_{-0.222}^{+0.196}$ & $0.729_{-0.217}^{+0.194}$ & $0.004$ & $0.002$ & $0.292_{-0.128}^{+0.163}$ & $0.209_{-0.918}^{+0.272}$ & $0.336$ & $0.083$ & $6.198_{-1.831}^{+2.151}$ & $-2.620_{-6.571}^{+9.608}$ & $0.451$ & $8.818$ \\
GW200129\_065458 & $0.891_{-0.246}^{+0.084}$ & $0.895_{-0.214}^{+0.078}$ & $0.005$ & $0.004$ & $0.079_{-0.150}^{+0.149}$ & $0.080_{-0.161}^{+0.159}$ & $0.005$ & $0.001$ & $-0.146_{-0.303}^{+0.314}$ & $-0.188_{-0.298}^{+0.383}$ & $0.015$ & $0.042$ \\
GW200311\_115853 & $0.707_{-0.331}^{+0.233}$ & $0.713_{-0.342}^{+0.226}$ & $0.006$ & $0.007$ & $-0.073_{-0.254}^{+0.208}$ & $-0.076_{-0.263}^{+0.200}$ & $0.006$ & $0.003$ & $-0.018_{-0.489}^{+0.408}$ & $-0.027_{-0.501}^{+0.415}$ & $0.006$ & $0.010$ \\
GW231206\_233901 & $0.546_{-0.380}^{+0.355}$ & $0.534_{-0.380}^{+0.354}$ & $0.007$ & $0.012$ & $-0.037_{-0.143}^{+0.125}$ & $-0.037_{-0.142}^{+0.115}$ & $0.008$ & $0.001$ & $-0.240_{-0.251}^{+0.229}$ & $-0.249_{-0.251}^{+0.232}$ & $0.006$ & $0.008$ \\
GW190412 & $0.252_{-0.137}^{+0.194}$ & $0.290_{-0.144}^{+0.210}$ & $0.030$ & $0.038$ & $0.286_{-0.123}^{+0.121}$ & $0.301_{-0.136}^{+0.139}$ & $0.016$ & $0.015$ & $0.026_{-0.127}^{+0.137}$ & $0.030_{-0.140}^{+0.151}$ & $0.007$ & $0.004$ \\
GW231028\_153006 & $0.573_{-0.296}^{+0.232}$ & $0.647_{-0.236}^{+0.199}$ & $0.057$ & $0.073$ & $0.439_{-0.119}^{+0.119}$ & $0.412_{-0.136}^{+0.135}$ & $0.034$ & $0.027$ & $0.003_{-6.970}^{+14.14}$ & $2.751_{-9.561}^{+12.00}$ & $0.058$ & $2.748$ \\
GW191109\_010717 & $0.537_{-0.337}^{+0.287}$ & $0.652_{-0.316}^{+0.237}$ & $0.069$ & $0.115$ & $-0.305_{-0.301}^{+0.283}$ & $-0.416_{-0.314}^{+0.361}$ & $0.051$ & $0.111$ & $4.005_{-12.36}^{+9.909}$ & $1.582_{-13.30}^{+12.39}$ & $0.031$ & $2.422$ \\
GW200224\_222234 & $0.545_{-0.387}^{+0.353}$ & $0.707_{-0.327}^{+0.236}$ & $0.107$ & $0.162$ & $0.114_{-0.170}^{+0.154}$ & $0.111_{-0.225}^{+0.172}$ & $0.013$ & $0.003$ & $0.157_{-0.355}^{+0.255}$ & $0.149_{-0.478}^{+0.293}$ & $0.010$ & $0.008$ \\
GW191204\_171526 & $0.447_{-0.288}^{+0.367}$ & $0.624_{-0.278}^{+0.270}$ & $0.143$ & $0.177$ & $0.144_{-0.121}^{+0.175}$ & $0.181_{-0.178}^{+0.599}$ & $0.100$ & $0.037$ & $-0.025_{-0.154}^{+0.191}$ & $0.019_{-0.212}^{+0.527}$ & $0.097$ & $0.044$ \\
GW230914\_111401 & $0.400_{-0.283}^{+0.412}$ & $0.629_{-0.307}^{+0.278}$ & $0.187$ & $0.229$ & $0.083_{-0.195}^{+0.164}$ & $0.096_{-0.241}^{+0.188}$ & $0.015$ & $0.013$ & $-0.441_{-11.82}^{+6.713}$ & $-0.480_{-11.82}^{+8.149}$ & $0.025$ & $0.039$ \\
GW170814 & $0.426_{-0.324}^{+0.402}$ & $0.655_{-0.313}^{+0.263}$ & $0.188$ & $0.229$ & $0.090_{-0.122}^{+0.130}$ & $0.098_{-0.153}^{+0.152}$ & $0.013$ & $0.008$ & $0.113_{-0.256}^{+0.231}$ & $0.130_{-0.262}^{+0.275}$ & $0.013$ & $0.017$ \\
GW200112\_155838 & $0.369_{-0.274}^{+0.427}$ & $0.626_{-0.308}^{+0.285}$ & $0.219$ & $0.258$ & $0.061_{-0.136}^{+0.147}$ & $0.077_{-0.187}^{+0.157}$ & $0.022$ & $0.016$ & $0.031_{-0.290}^{+0.285}$ & $0.016_{-0.295}^{+0.280}$ & $0.007$ & $0.015$ \\
GW231226\_101520 & $0.364_{-0.256}^{+0.370}$ & $0.609_{-0.284}^{+0.279}$ & $0.244$ & $0.245$ & $-0.091_{-0.105}^{+0.084}$ & $-0.131_{-0.136}^{+0.113}$ & $0.076$ & $0.040$ & $0.024_{-0.134}^{+0.137}$ & $-0.011_{-0.154}^{+0.158}$ & $0.043$ & $0.035$ \\
GW150914 & $0.309_{-0.239}^{+0.433}$ & $0.605_{-0.296}^{+0.280}$ & $0.294$ & $0.296$ & $-0.034_{-0.142}^{+0.111}$ & $-0.066_{-0.201}^{+0.149}$ & $0.047$ & $0.031$ & $0.178_{-0.177}^{+0.197}$ & $0.172_{-0.204}^{+0.203}$ & $0.010$ & $0.006$ \\
GW230927\_153832 & $0.403_{-0.285}^{+0.367}$ & $0.674_{-0.274}^{+0.246}$ & $0.295$ & $0.271$ & $0.071_{-0.102}^{+0.116}$ & $0.067_{-0.143}^{+0.138}$ & $0.025$ & $0.004$ & $0.058_{-0.163}^{+0.155}$ & $0.052_{-0.190}^{+0.184}$ & $0.018$ & $0.006$ \\
GW191216\_213338 & $0.284_{-0.205}^{+0.368}$ & $0.608_{-0.376}^{+0.272}$ & $0.298$ & $0.324$ & $0.148_{-0.146}^{+0.211}$ & $0.268_{-0.224}^{+0.408}$ & $0.152$ & $0.120$ & $0.035_{-0.195}^{+0.223}$ & $0.194_{-0.238}^{+0.293}$ & $0.205$ & $0.159$ \\
GW230814\_230901 & $0.160_{-0.121}^{+0.226}$ & $0.598_{-0.254}^{+0.192}$ & $0.769$ & $0.438$ & $-0.022_{-0.061}^{+0.046}$ & $-0.037_{-0.089}^{+0.086}$ & $0.097$ & $0.016$ & $0.032_{-0.088}^{+0.090}$ & $0.036_{-0.097}^{+0.093}$ & $0.008$ & $0.003$ \\
GW250114 & $0.164_{-0.131}^{+0.240}$ & $0.704_{-0.268}^{+0.221}$ & $0.859$ & $0.539$ & $-0.032_{-0.044}^{+0.034}$ & $-0.072_{-0.055}^{+0.056}$ & $0.282$ & $0.040$ & $0.001_{-0.050}^{+0.049}$ & $-0.026_{-0.081}^{+0.065}$ & $0.124$ & $0.027$ \\
\noalign{\hrule height 0.8pt}
\end{tabular}%
}
\end{table*}
We find that the posterior distributions of $\chi_p$ exhibit significantly larger variations than those of $\chi_{\rm eff}$ under different spin priors. In particular, $\chi_p$ shows systematically higher JSD and more pronounced median shifts across most events, indicating a strong sensitivity to the choice of prior~\cite{Prasad:2026idz,Talbot:2018}. In contrast, the posterior distributions of $\chi_{\rm eff}$ remain comparatively stable, with smaller changes in both distribution shape and central value. This result quantitatively indicates that $\chi_{\rm eff}$ is less dependent on the choice of prior and exhibits greater stability than $\chi_p$~\cite{Vitale:2017,LIGOScientific:2021psn}.

For $\delta \hat{\phi}_3$, the JSD values are generally small, indicating overall robustness with respect to the choice of spin prior. Notable exceptions include GW231123\_135430, which exhibits the largest JSD in the sample and also shows the most significant median shift, reaching $\sim 8.8$. However, parameter estimation for GW231123\_135430 has been shown to be particularly sensitive to the choice of waveform model, with substantial discrepancies reported among different waveform families~\cite{LIGOScientific:2025GW231123,Jan:2026GW231123,Ray:2025GW231123,Hu:2026GW231123}. Results strongly influenced by this event should therefore be interpreted with caution. In addition, GW250114 and GW191216\_213338 also display relatively large JSD values compared to the rest of the sample. In contrast, for short-inspiral events such as GW231028\_153006 and GW191109\_010717, the JSD values remain small, while the corresponding median shifts are noticeably larger, reflecting the limited information content of short inspiral signals.~\cite{Seymour:2026bjg,Poisson:1995}.

These results indicate that changes in the overall distribution shape do not necessarily correspond to shifts in central tendency, and thus cannot fully characterize the impact of the prior on the posterior~\cite{Lin:1991zzm,Endres:2003}. This behavior is also consistent with the presence of parameter degeneracies between spin effects and PN-deviation parameters at similar orders in the waveform phasing~\cite{Cutler:1994,Narikawa:2016deg}. A comprehensive assessment therefore requires a multi-dimensional comparison, combining both distribution-level metrics (e.g., JSD) and point-estimate shifts (e.g., $|\Delta \mathrm{med}|$).

\subsection{Correlation between spin parameters and $\delta \hat{\phi}_3$}

We next investigate the relationship between the prior sensitivity of spin parameters and that of the deviation parameter $\delta \hat{\phi}_3$, which enters at the same 1.5PN order~\cite{Blanchet:2013haa,Narikawa:2016deg}. To quantify prior sensitivity, we use the JSD as a measure of the difference between posterior distributions obtained under different spin priors~\cite{Lin:1991zzm,Endres:2003}.

\begin{figure}[t]
    \centering
    \includegraphics[width=0.65\linewidth]{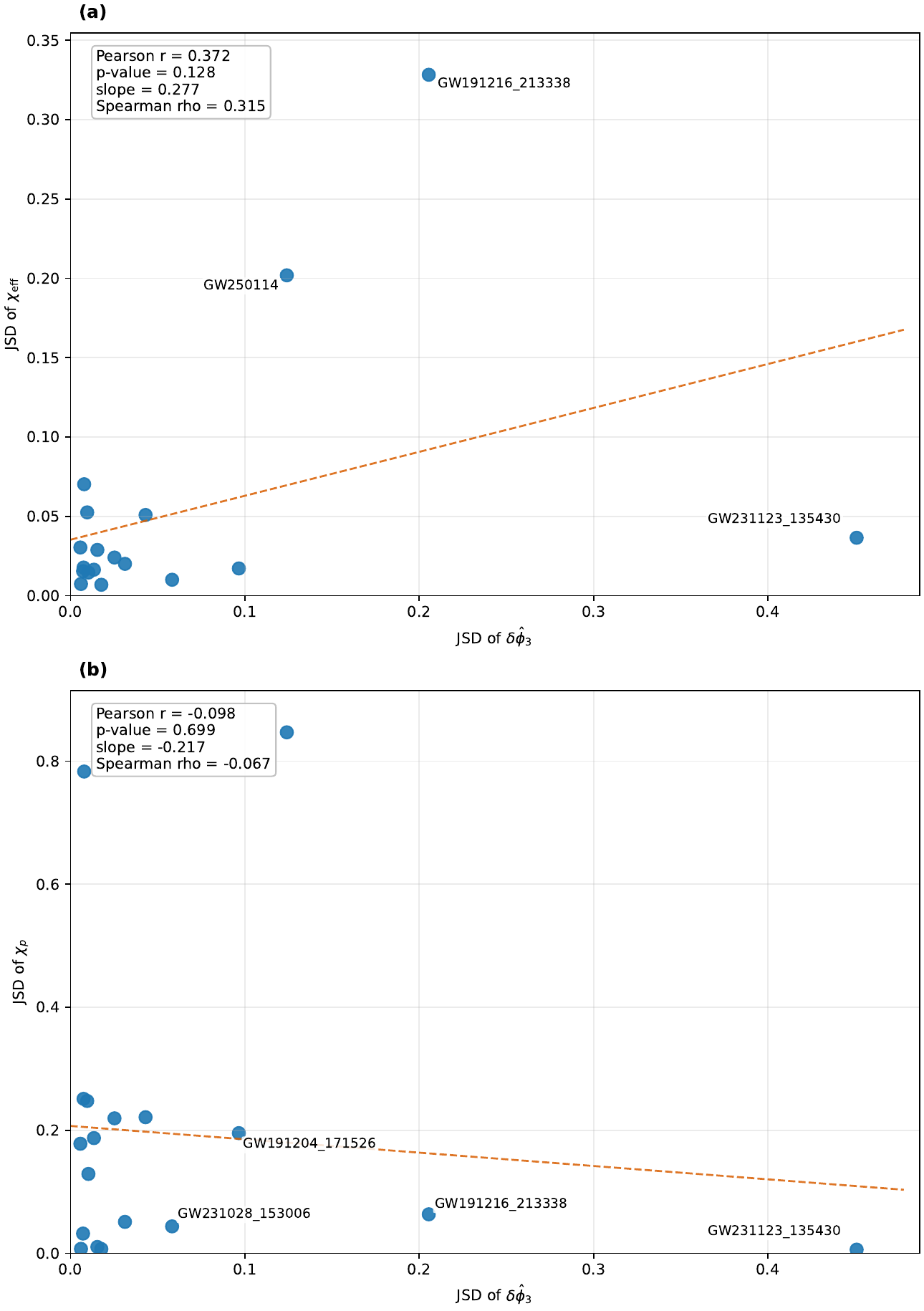}
    \caption{
Correlation between the JSD of $\delta \hat{\phi}_3$ and those of $\chi_{\rm eff}$ (a) and $\chi_p$ (b). Each point represents one event. JSD measures posterior differences under different spin priors. Dashed lines show linear fits; Pearson $r$, $p$-value, slope, and Spearman $\rho$ quantify the correlations.
}
    \label{fig:gr}
\end{figure}

The results are shown in FIG.~\hyperref[fig:gr]{3}, where we compare the JSD of $\delta \hat{\phi}_3$ with those of the spin parameters. The top panel presents the correlation between the JSDs of $\chi_{\rm eff}$ and $\delta \hat{\phi}_3$, while the bottom panel shows the corresponding comparison for $\chi_p$. We emphasize that the JSDs of $\chi_{\rm eff}$ and $\chi_p$ are computed from parameter-estimation runs that do not include $\delta \hat{\phi}_3$, whereas the JSD of $\delta \hat{\phi}_3$ is obtained from analyses in which the deviation parameter is included. This setup allows us to assess whether the prior sensitivity of spin parameters inferred within the standard GR framework exhibits any correlation or coupling with the variation of $\delta \hat{\phi}_3$~\cite{Cutler:1994,Seymour:2026bjg}.

We quantify the degree of correlation using four statistical measures: the Pearson correlation coefficient $r$, the associated $p$-value, the slope of the best-fit linear relation, and the Spearman rank correlation coefficient $\rho$~\cite{Copparoni:2025vty}.
We find that when the JSDs of $\chi_{\rm eff}$ and $\chi_p$ are computed from standard GR parameter-estimation runs, they exhibit no statistically significant linear correlation with the JSD of $\delta \hat{\phi}_3$ in FIG.~\hyperref[fig:gr]{3}. Most events cluster near the origin, indicating that both the spin parameters and $\delta \hat{\phi}_3$ show weak prior sensitivity in these cases~\cite{Prasad:2026idz,Talbot:2018}. In contrast, a subset of events lies significantly farther from the origin, corresponding to cases with stronger prior sensitivity. These events are explicitly labeled in the figure, regardless of their alignment with the best-fit linear relation, to emphasize the variation in prior sensitivity across different events.

\begin{figure}[t]
    \centering
    \includegraphics[width=0.65\linewidth]{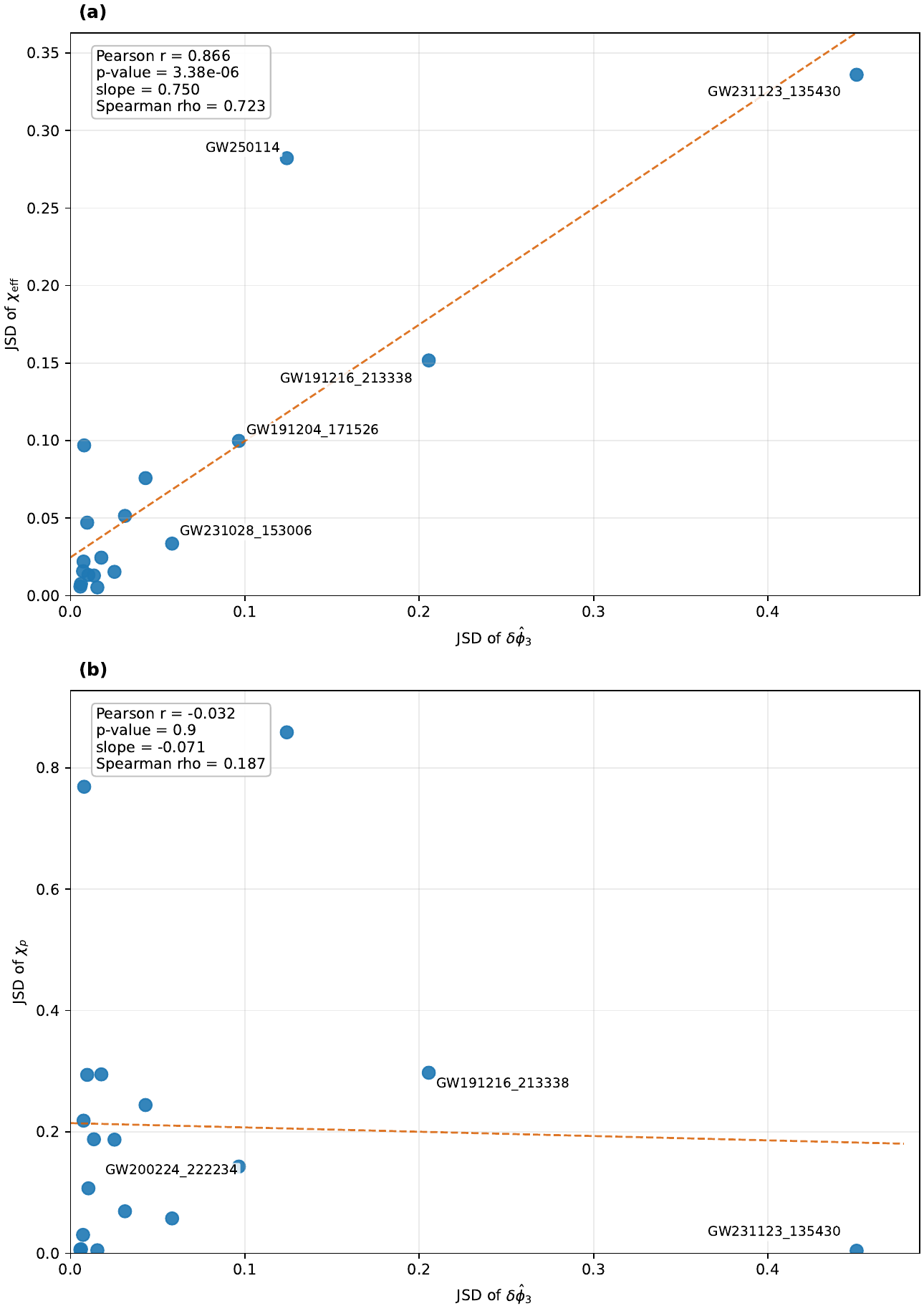}
    \caption{
Same as Fig.~\ref{fig:gr}, but with all JSDs computed from analyses including $\delta \hat{\phi}_3$. A strong correlation is observed with $\chi_{\rm eff}$ (a), while the correlation with $\chi_p$ (b) remains weaker.
}
    \label{fig:non_gr}
\end{figure}

Using the same setup, we obtain Fig.~\hyperref[fig:non_gr]{4}. The key difference from Fig.~\hyperref[fig:gr]{3} is that the JSDs of the spin parameters are now computed from parameter-estimation runs that include the deviation parameter $\delta \hat{\phi}_3$, i.e., using posterior samples generated with both $\delta \hat{\phi}_3$ and the modified spin prior. For detailed comparisons of the changes in $\chi_{\rm eff}$ and $\chi_p$, we refer the reader to the Appendix.

In contrast to the results obtained in the standard GR framework, we now observe a clear change in the correlation behavior. While $\chi_p$ still does not exhibit a statistically significant linear correlation with $\delta \hat{\phi}_3$ (Pearson $r = -0.032$, $p = 0.071$), $\chi_{\rm eff}$ shows a strong and statistically significant linear correlation (Pearson $r = 0.866$, $p = 3.38 \times 10^{-6}$).

A possible explanation for this behavior is that, within the non-GR parameter-estimation framework, both the spin parameters and the deviation parameter $\delta \hat{\phi}_3$ are inferred simultaneously and can share common phase information~\cite{Cutler:1994,Seymour:2026bjg}. As a result, variations in the prior may affect their posteriors in a correlated manner~\cite{Prasad:2026idz,Talbot:2018}. In particular, since both $\chi_{\rm eff}$ and $\delta \hat{\phi}_3$ primarily influence the waveform phase at the 1.5PN order, a partial degeneracy between these parameters may arise, leading to an apparent statistical correlation~\cite{Narikawa:2016deg,Seymour:2026bjg}. This provides a natural interpretation for the stronger correlation observed with $\chi_{\rm eff}$, while the weaker correlation with $\chi_p$ suggests that precession effects are less directly coupled to the deviation parameter~\cite{Petersen:2014exq,Vitale:2017}.
Notably, the event GW231123\_135430 appears to be particularly sensitive in this relation, lying at the high end of both JSDs and contributing strongly to the observed trend.

To further assess the robustness of this correlation, we perform a leave-one-out test for Fig.~\hyperref[fig]{4}(a), in order to evaluate and mitigate the impact of particular events such as GW231123\_135430 on the overall result. Specifically, we remove each event in turn from the sample of 18 events and evaluate the resulting change in the linear correlation between $\chi_{\rm eff}$ and $\delta \hat{\phi}_3$, in order to determine whether the observed trend is driven by a small number of events. We use a single metric, the Pearson correlation coefficient $r$ and its corresponding $p$-value, to quantify the strength and robustness of the linear correlation. The maximum and minimum cases are finally selected, and the corresponding results are summarized in TABLE~\hyperref[tab:loo_extremes]{II}.
\begin{table}[t]
\centering
\caption{Extreme values of the Pearson correlation coefficient between event-wise JSDs of $\chi_{\rm eff}$ and $\delta\hat{\phi}_3$, with corresponding $p$-values obtained from the leave-one-out test.}
\label{tab:loo_extremes}

\begin{tabular}{lccc}
\noalign{\hrule height 0.8pt}
Removed event & Pearson $r$ & $p$-value & Description \\
\noalign{\hrule height 0.5pt}
GW250114 & 0.955 & $2.7 \times 10^{-9}$ & Maximum correlation \\
GW231123\_135430 & 0.752 & $5.0 \times 10^{-4}$ & Minimum correlation \\
\noalign{\hrule height 0.8pt}
\end{tabular}

\end{table}

 We find that removing GW250114 leads to the strongest correlation, with $r = 0.955$ and a highly significant $p$-value of $2.7 \times 10^{-9}$, indicating that this event reduces the overall correlation when included. In contrast, removing GW231123\_135430 yields the weakest correlation, with $r = 0.752$ and a larger $p$-value of $5.0 \times 10^{-4}$, indicating that this event has a substantial influence on the overall correlation strength. Its inferred properties are known to be sensitive to waveform modeling and possible event-specific systematics. Nevertheless, the correlation persists when this event is removed.

These results demonstrate that the correlation is sensitive to the inclusion of specific events and is not uniformly supported across the full sample, but instead partially driven by a subset of high-impact events with stronger prior sensitivity~\cite{Prasad:2026idz,Talbot:2018}.
However, this does not invalidate the previous interpretation. Rather, it indicates that the observed correlation may be influenced by the limited sample size and the presence of a few high-impact events. Future observations with a larger number of events, particularly those exhibiting strong sensitivity to spin priors, will be essential to further assess the robustness and physical origin of this correlation.

\section{Conclusion}

In this work, we have investigated the impact of spin priors on parameterized tests of General Relativity, focusing on the 1.5PN deviation parameter $\delta \hat{\phi}_3$. Using real gravitational-wave events, we quantified prior-induced effects through the JSD and median shifts of posterior distributions.
We find that the effective precession spin parameter $\chi_p$ is significantly more sensitive to the choice of spin prior than the effective inspiral spin parameter $\chi_{\rm eff}$, whose posterior remains comparatively stable. For the deviation parameter $\delta \hat{\phi}_3$, the overall JSD values are generally small, indicating a degree of robustness. However, in low-information regimes, large shifts in the posterior median can still occur, demonstrating that distribution shape alone does not fully capture prior-induced effects. It should be noted that the inferred properties of GW231123\_135430 are sensitive to waveform modeling and possible event-specific systematics. Results strongly influenced by this event should therefore be interpreted with caution.

By analyzing the correlation between the prior sensitivity of spin parameters and that of $\delta \hat{\phi}_3$, we show that no significant correlation is present when spin parameters are inferred within the standard GR framework. In contrast, when $\delta \hat{\phi}_3$ is included in the parameter-estimation model, a strong correlation emerges between $\chi_{\rm eff}$ and $\delta \hat{\phi}_3$, while $\chi_p$ remains weakly correlated. This behavior can be naturally explained by a partial degeneracy between $\chi_{\rm eff}$ and $\delta \hat{\phi}_3$, as both parameters contribute to the waveform phase at the 1.5PN order. A leave-one-out test further reveals that the observed correlation is sensitive to the inclusion of specific events, particularly GW231123\_135430. Removing this event leads to a weaker correlation, although the correlation persists, indicating that GW231123\_135430 has a substantial influence on the overall correlation strength without driving the result entirely.

Several limitations of the present work should be noted. First, although our results from real events reveal suggestive trends, a more systematic injection campaign will be necessary to fully disentangle the roles of prior choice, parameter degeneracy, signal-to-noise ratio, and spin information in shaping the posterior behavior of $\delta \hat{\phi}_3$. Such controlled injection studies will provide a cleaner framework for testing the robustness of the correlations reported here.

Second, our analysis focuses on a single deviation parameter, $\delta \hat{\phi}_3$. In realistic gravitational-wave signals, however, possible departures from General Relativity may not be confined to a single parameter, but could instead arise from the combined effect of multiple deviation parameters. While the single-parameter approach adopted here is useful for isolating the propagation of prior-induced bias, extending the analysis to a higher-dimensional non-GR parameter space will be important for assessing the generality of our conclusions.

Overall, our results highlight that prior-induced biases in spin parameters can propagate into parameterized tests of General Relativity in a non-trivial and model-dependent manner. This emphasizes the importance of carefully accounting for prior choices in gravitational-wave inference. Future observations with larger event samples, particularly those with strong spin sensitivity, together with controlled injection studies, will be crucial for assessing the robustness and physical origin of these effects.

\section*{Acknowledgements}
We are grateful to  Tao Zhu and Hactor Villarrubia-Rojo for their helpful discussions. We also thank Gregory Ashton for his assistance with the \texttt{bilby} package. This work was supported by the National Natural Science Foundation of China under Grant No.
12065007.

\appendix

\section*{Appendix}
\label{app:additional}
\begin{figure}
    \centering
    \includegraphics[width=\textwidth]{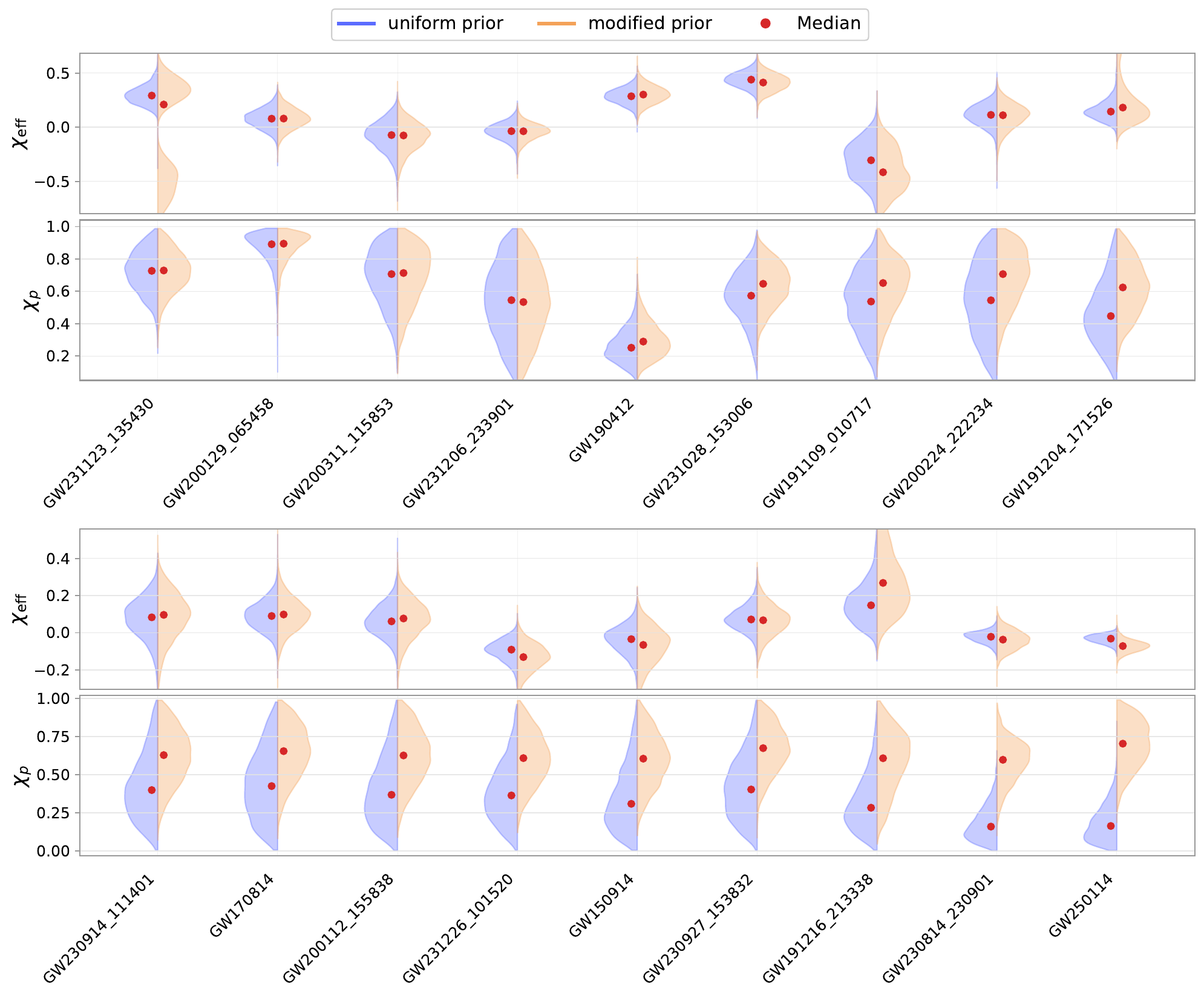}
   \caption{Deviations of $\chi_{\rm eff}$ and $\chi_p$ under waveform models including deviation parameters. The posterior distributions and their associated biases exhibit significant differences compared to \hyperref[fig:fig1]{Fig.~1}.}
    \label{fig:spin2}
\end{figure}

In this appendix, we present supplementary results for the posterior distributions of $\chi_p$ and $\chi_{\rm eff}$, together with their corresponding Jensen--Shannon divergence (JSD), obtained from waveform models including the deviation parameter $\delta \hat{\phi}_3$. For several events, the inferred posteriors show significant deviations from those obtained under the standard GR assumption. Detailed results are provided in Fig.~\ref{fig:spin2} and Table~\ref{tab:spin-jsd-delta-summary}.

\begin{table*}[t]
\centering
\large
\renewcommand{\arraystretch}{2}
\setlength{\tabcolsep}{3.2pt}

\caption{Posterior medians with symmetric 90\% credible intervals, Jensen--Shannon divergence (JSD), absolute median shifts $|\Delta \mathrm{med}|$, and absolute JSD differences $|\Delta \mathrm{JSD}|$ for $\chi_p$ and $\chi_{\rm eff}$.}
\label{tab:spin-jsd-delta-summary}

\resizebox{\textwidth}{!}{%
\begin{tabular}{l|ccccc|ccccc}
\hline
\multirow{2}{*}{{\fontsize{20}{22}\selectfont\textbf{Events}}}
& \multicolumn{5}{c|}{{\fontsize{18}{18}\selectfont$\chi_p$}}
& \multicolumn{5}{c}{{\fontsize{18}{18}\selectfont$\chi_{\rm eff}$}} \\
\cline{2-11}
& {\fontsize{13}{16}\selectfont uniform prior}
& {\fontsize{13}{16}\selectfont modified prior}
& {\fontsize{13}{16}\selectfont JSD}
& {\fontsize{13}{16}\selectfont $|\Delta {\rm med}|$}
& {\fontsize{13}{16}\selectfont $|\Delta {\rm JSD}|$}
& {\fontsize{13}{16}\selectfont uniform prior}
& {\fontsize{13}{16}\selectfont modified prior}
& {\fontsize{13}{16}\selectfont JSD}
& {\fontsize{13}{16}\selectfont $|\Delta {\rm med}|$}
& {\fontsize{13}{16}\selectfont $|\Delta {\rm JSD}|$} \\
\hline
GW231123\_135430 & $0.698_{-0.239}^{+0.165}$ & $0.69_{-0.259}^{+0.179}$ & $0.006$ & $0.008$ & $0.002$ & $-0.369_{-0.224}^{+0.183}$ & $-0.402_{-0.28}^{+0.246}$ & $0.037$ & $0.033$ & $0.299$ \\
GW230927\_153832 & $0.373_{-0.283}^{+0.384}$ & $0.395_{-0.298}^{+0.409}$ & $0.007$ & $0.022$ & $0.288$ & $0.039_{-0.071}^{+0.07}$ & $0.042_{-0.073}^{+0.074}$ & $0.007$ & $0.004$ & $0.018$ \\
GW231206\_233901 & $0.539_{-0.391}^{+0.351}$ & $0.552_{-0.385}^{+0.349}$ & $0.007$ & $0.013$ & $0$ & $-0.038_{-0.17}^{+0.135}$ & $-0.046_{-0.167}^{+0.139}$ & $0.007$ & $0.009$ & $0$ \\
GW200129\_065458 & $0.89_{-0.256}^{+0.084}$ & $0.901_{-0.185}^{+0.073}$ & $0.01$ & $0.011$ & $0.005$ & $0.094_{-0.167}^{+0.162}$ & $0.134_{-0.207}^{+0.146}$ & $0.029$ & $0.04$ & $0.024$ \\
GW190412 & $0.25_{-0.135}^{+0.189}$ & $0.291_{-0.145}^{+0.21}$ & $0.032$ & $0.041$ & $0.002$ & $0.267_{-0.096}^{+0.106}$ & $0.28_{-0.109}^{+0.114}$ & $0.015$ & $0.013$ & $0$ \\
GW231028\_153006 & $0.498_{-0.292}^{+0.306}$ & $0.577_{-0.271}^{+0.263}$ & $0.044$ & $0.079$ & $0.014$ & $0.473_{-0.143}^{+0.128}$ & $0.484_{-0.159}^{+0.134}$ & $0.01$ & $0.011$ & $0.024$ \\
GW191109\_010717 & $0.489_{-0.309}^{+0.319}$ & $0.588_{-0.297}^{+0.264}$ & $0.051$ & $0.099$ & $0.018$ & $-0.384_{-0.243}^{+0.275}$ & $-0.425_{-0.271}^{+0.307}$ & $0.02$ & $0.041$ & $0.031$ \\
GW191216\_213338 & $0.253_{-0.17}^{+0.354}$ & $0.312_{-0.221}^{+0.526}$ & $0.063$ & $0.058$ & $0.234$ & $0.106_{-0.061}^{+0.118}$ & $0.218_{-0.19}^{+0.406}$ & $0.328$ & $0.112$ & $0.177$ \\
GW200224\_222234 & $0.539_{-0.391}^{+0.365}$ & $0.721_{-0.336}^{+0.219}$ & $0.129$ & $0.183$ & $0.022$ & $0.102_{-0.162}^{+0.135}$ & $0.097_{-0.206}^{+0.154}$ & $0.015$ & $0.005$ & $0.001$ \\
GW200311\_115853 & $0.466_{-0.36}^{+0.405}$ & $0.7_{-0.325}^{+0.238}$ & $0.178$ & $0.234$ & $0.172$ & $-0.047_{-0.197}^{+0.148}$ & $-0.074_{-0.251}^{+0.198}$ & $0.03$ & $0.027$ & $0.024$ \\
GW170814 & $0.411_{-0.319}^{+0.431}$ & $0.658_{-0.324}^{+0.257}$ & $0.187$ & $0.247$ & $0.001$ & $0.076_{-0.12}^{+0.119}$ & $0.09_{-0.145}^{+0.129}$ & $0.016$ & $0.014$ & $0.003$ \\
GW191204\_171526 & $0.427_{-0.282}^{+0.357}$ & $0.64_{-0.276}^{+0.27}$ & $0.195$ & $0.213$ & $0.052$ & $0.157_{-0.044}^{+0.08}$ & $0.159_{-0.056}^{+0.077}$ & $0.017$ & $0.003$ & $0.083$ \\
GW230914\_111401 & $0.351_{-0.25}^{+0.383}$ & $0.584_{-0.292}^{+0.31}$ & $0.219$ & $0.234$ & $0.032$ & $0.122_{-0.194}^{+0.178}$ & $0.154_{-0.256}^{+0.185}$ & $0.024$ & $0.032$ & $0.009$ \\
GW231226\_101520 & $0.355_{-0.247}^{+0.329}$ & $0.574_{-0.277}^{+0.274}$ & $0.221$ & $0.219$ & $0.023$ & $-0.089_{-0.098}^{+0.078}$ & $-0.118_{-0.112}^{+0.088}$ & $0.051$ & $0.029$ & $0.025$ \\
GW150914 & $0.414_{-0.326}^{+0.403}$ & $0.682_{-0.302}^{+0.243}$ & $0.248$ & $0.268$ & $0.046$ & $-0.015_{-0.126}^{+0.104}$ & $-0.049_{-0.17}^{+0.142}$ & $0.053$ & $0.034$ & $0.005$ \\
GW200112\_155838 & $0.367_{-0.277}^{+0.401}$ & $0.644_{-0.313}^{+0.274}$ & $0.251$ & $0.277$ & $0.032$ & $0.063_{-0.145}^{+0.143}$ & $0.079_{-0.184}^{+0.151}$ & $0.018$ & $0.017$ & $0.004$ \\
GW230814\_230901 & $0.113_{-0.087}^{+0.215}$ & $0.518_{-0.215}^{+0.213}$ & $0.783$ & $0.405$ & $0.014$ & $-0.016_{-0.052}^{+0.042}$ & $-0.015_{-0.076}^{+0.074}$ & $0.07$ & $0$ & $0.027$ \\
GW250114 & $0.157_{-0.128}^{+0.239}$ & $0.691_{-0.267}^{+0.235}$ & $0.847$ & $0.535$ & $0.012$ & $-0.031_{-0.044}^{+0.035}$ & $-0.061_{-0.05}^{+0.072}$ & $0.202$ & $0.03$ & $0.08$ \\
\noalign{\hrule height 0.8pt}
\end{tabular}%
}
\end{table*}

\bibliographystyle{apsrev4-1}
\bibliography{refs}
\clearpage
\end{document}